\documentclass[runningheads]{llncs}
\usepackage[T1]{fontenc}
\usepackage{xspace}
\usepackage{amsmath,amssymb,amsfonts}
\usepackage{graphicx}
\usepackage{textcomp}
\usepackage{xcolor}
\usepackage{soul}
\usepackage{xspace}
\usepackage{url}
\usepackage{todonotes}
\usepackage{enumitem}
\usepackage{colortbl} 
\usepackage{hyperref}
\usepackage{cleveref}
\usepackage{listings} 
\usepackage{graphicx}
\usepackage{subcaption}
\usepackage{todonotes}
\usepackage{tikz}
\usetikzlibrary{arrows.meta, positioning, shapes.multipart}

\def\BibTeX{{\rm B\kern-.05em{\sc i\kern-.025em b}\kern-.08em
    T\kern-.1667em\lower.7ex\hbox{E}\kern-.125emX}}

\crefformat{section}{§#2#1#3}
\Crefformat{section}{§#2#1#3}

\newcommand{\tool}{\textsc{GreenMalloc}\xspace}
\newcommand{\googlemem}{\textsc{TCMalloc}\xspace}
\newcommand{\gnumem}{\textsc{glibc malloc}\xspace}
\newcommand{\gem}{\textsc{gem5}\xspace}
\newcommand{\searchsys}{\textsc{SearchSYS}\xspace}
\newcommand{\randmalloc}{\textsc{rand\_malloc}\xspace}

\author{Aidan Dakhama\inst{1}\orcidID{0009-0002-7318-7964} \and
W.B. Langdon\inst{2}\orcidID{0000-0002-6388-4160} \and
Hector D. Menendez\inst{1}\orcidID{0000-0002-6314-3725} \and 
\\ Karine Even-Mendoza\inst{1}\orcidID{0000-0002-3099-1189}}

\authorrunning{Dakhama et al.}

\institute{King’s College London, London, UK. 
\email{\{aidan.dakhama, karine.even\_mendoza, hector.menendez\}@kcl.ac.uk} \and
University College London, London, UK. 
\email{w.langdon@ucl.ac.uk}}

\begin{document}
\title{GreenMalloc: Allocator Optimisation for Industrial Workloads}
%
%
%
\authorrunning{Aidan Dakhama {\em et al.}}
%
%
\maketitle              
\begin{abstract}
We present \tool{}, a multi-objective search-based framework for automatically configuring memory allocators. Our approach uses \mbox{NSGA-II} and \randmalloc{} as a lightweight proxy benchmarking tool, we efficiently explore allocator parameters from execution traces and transfer the best configurations to \gem{}, a large system simulator, in a case study on two allocators: the GNU C/C++ compiler's \gnumem{} and Google's \googlemem{}.
Across diverse workloads, our empirical results show up to 4.1\% reduction in average heap usage 
without loss of runtime efficiency,
indeed we get a 0.25\% reduction.

\end{abstract}

\section{Introduction}

Efficient memory management remains a challenge in modern computing, with empirical studies showing that allocator choice and configuration significantly affect performance and resource use~\cite{AllocComp2021,MemAllocWarehouse24}, and consequently influence energy consumption \cite{MemVsGreen2017}. 
Memory allocators such as the GNU allocator (\gnumem)~\cite{glibc_manual} and Google’s \googlemem{}~\cite{google_tcmalloc} are widely deployed, but tuning their run-time parameters, which allows fine-grained control of their operation, is difficult since optimal settings vary substantially with workload and system behaviour. As a result, production systems often rely on default configurations, which can waste memory, increase energy use, and degrade performance.


The \gem{} simulator has become a de facto standard for hardware–software co-design evaluation in academia and industry~\cite{binkert2011gem5}. \gem{} is a large, complex codebase with unusual allocation patterns. Simulations are notoriously slow, often lasting minutes to days, magnifying the impact of allocator efficiency while making manual parameter tuning infeasible. Even modest improvements in heap usage or allocation overhead can shorten runtimes, reduce computational costs, and lower the energy footprint of repeated experiments. Among the many potential optimisation targets, allocator parameters represent a promising but underexplored opportunity to improve runtime and memory consumption in \gem{}. Yet, allocator optimisation in realistic workloads is challenging: exploration is costly due to long execution times and high memory demands, and results shall be relevant to system behaviour in the wild.


We show it is possible to automatically tune memory allocator parameters to reduce heap usage and energy consumption in industrial workloads.  By targeting both memory and runtime, we identify allocator configurations that yield practical improvements across different contexts.
We employ a search-based optimisation approach to tune allocator parameters and evaluate their impact on heap usage and runtime in \gem{}’s System Emulation (SE) mode~\cite{binkert2011gem5}. To avoid the prohibitive cost of optimising directly on full simulations, we use \randmalloc{}~\cite{Langdon:2025:UKCI} as a lightweight proxy benchmark, enabling efficient exploration of the parameter space before deploying promising configurations in \gem{}.


Experimental results demonstrate that while peak heap usage remained largely unchanged, 
\gnumem{} tuning yields consistent reductions in average memory ($\approx$4\% improvement), a big improvement in free rate ($\approx$2.4× faster release of memory), and a small but measurable reduction in instructions ($\approx$0.25\% less). \mbox{\googlemem{}} shows more modest improvements in stability and predictability, with small gains in memory free rate and instruction count.
These findings suggest that automated allocator tuning significantly reduces computational and energy footprints of long-running industrial simulations keeping reliability and performance.
This paper makes the following key contributions:
\begin{itemize}[nosep,noitemsep,leftmargin=11pt,itemsep=0pt, topsep=0pt]
    \item A novel search-based optimisation methodology for configuring memory allocators to improve performance and energy efficiency.
    
    \item \tool{}, a prototype implementation of this approach, designed to be generalisable beyond the allocators studied in this paper.  
    
    \item A systematic study of allocator tuning using \tool{}, applied to two widely used allocators, \gnumem{} and \googlemem{}, in the context of \gem{}, a complex and industrially relevant system.  
    
\end{itemize}

\vspace{0.2cm}
\noindent\textit{Availability. }
\tool{} and all artifacts are at \href{https://zenodo.org/records/17182847#:~:text=Files-,README,-.pdf}{\raisebox{-0.4ex}{\includegraphics[height=1.1em]{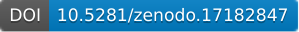}}}.
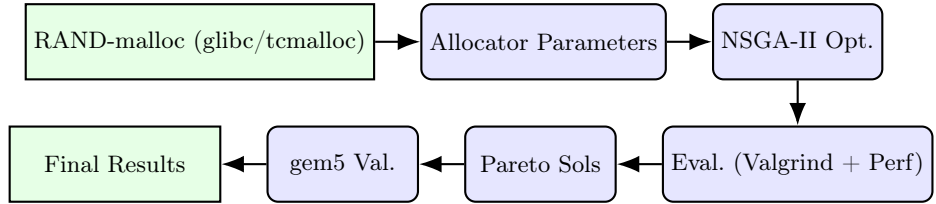
\begin{figure}[t!]
    \centering
\begin{tikzpicture}[
    node distance=0.6cm and 0.6cm,
    process/.style={rectangle, rounded corners, draw=black, thick, fill=blue!10, text centered, minimum width=2cm, minimum height=1cm},
    io/.style={rectangle, draw=black, thick, fill=green!10, text centered, minimum width=2.8cm, minimum height=1cm},
    decision/.style={diamond, draw=black, thick, fill=orange!10, aspect=2, text centered, inner sep=1pt},
    arrow/.style={-{Latex[length=3mm]}, thick},
]

\node[io] (randmalloc) {RAND-malloc (glibc/tcmalloc)};
\node[process, right=of randmalloc] (params) {Allocator Parameters};
\node[process, right=of params] (nsga) {NSGA-II Opt.};
\node[process, below=of nsga] (eval) {Eval.  (Valgrind + Perf)};
\node[process, left=of eval] (pareto) {Pareto Sols};
\node[process, left=of pareto] (gem5) {gem5 Val.};
\node[io, left=of gem5] (results) {Final Results};

\draw[arrow] (randmalloc) -- (params);
\draw[arrow] (params) -- (nsga);
\draw[arrow] (nsga) -- (eval);
\draw[arrow] (eval) -- (pareto);
\draw[arrow] (pareto) -- (gem5);
\draw[arrow] (gem5) -- (results);

\end{tikzpicture}
    \vspace{-0.5 cm}
    \caption{General \tool{} workflow: starting with \randmalloc{} optimisation to identify efficient allocation parameters, ended by validation on \gem{} to assess improvements in memory usage and runtime.}
    \label{fig:diagram}
    \vspace{-0.5cm}
\end{figure}

\section{Methodology}

\autoref{fig:diagram} shows the whole system workflow. First, we optimise over a synthetic benchmark \randmalloc{}~\cite{Langdon:2025:UKCI} to explore the parameter search space efficiently; next, we run the resulting optimal parameters on \gem{} simulator to evaluate the transferability to the real system under complex running conditions.

\subsection{Heap Memory Allocator Parameters}

Both allocators expose several tunable parameters that govern allocation behaviour. For \gnumem{}, parameters including thresholds for switching between \texttt{mmap} and \texttt{sbrk} allocation, trimming and padding behaviour, and arena limits, all of which were extracted from the \gnumem{} manual~\cite{glibc_manual}. For \googlemem{}, there is a range of tunable parameters, 
such as release rates, thread cache size, page size overrides, and heap limits and were extracted from Google's \googlemem{} documentation~\cite{gperftools_tcmalloc}. These parameters form a high-dimensional, mixed discrete-continuous search space with non-trivial interactions, making manual tuning impractical. A summary of these parameters is in our artifact \cite{zenodo:green}.

\subsection{Genetic Algorithm Optimisation with \texttt{pymoo}}

We employ the genetic algorithm (GA), \texttt{NSGA-II}, implemented using \texttt{pymoo}, selected for its effectiveness in multi-objective optimisation. Each candidate encodes a \textit{configuration of allocator parameters}, which are passed as environment variables to the allocator implementation (\texttt{glibc malloc} or \texttt{tcmalloc}). Standard GA operators (mutation, crossover, and elitism) are applied to evolve the population towards better-performing configurations.

\vspace{0.1 cm}
\noindent We formulate the optimisation as a multi-objective problem, jointly targeting \emph{peak heap usage} and \emph{execution time} to balance sustainability and performance: 
\begin{itemize}[nosep,noitemsep,leftmargin=11pt,itemsep=0pt, topsep=0pt]
\item \textit{Green Allocation. }%
Peak heap usage, measured using \texttt{valgrind}'s \texttt{massif} tool to minimise the total memory consumed by \gem{} at any given point. Lowering peak memory has two sustainability benefits: it reduces hardware requirements, and enables better workload co-location. Whilst average heap usage may correlate with runtime energy consumption, peak heap determines the minimal system configuration required. Our evaluation measures both metrics.

\item \textit{Performance. }%
Execution time, measured with the \texttt{time} utility to balance memory efficiency against performance. This ensures that improvements do not come at the cost of excessively slow executions. 
\end{itemize}
For each generation, \texttt{NSGA-II} evaluates candidate parameter vectors by executing \randmalloc{} and recording the two optimisation objectives. This yields a Pareto front of non-dominated solutions balancing memory usage and runtime. 
From this front, we select representative candidates: the one minimising runtime, the one minimising memory, and a balanced solution near the centre of the trade-off curve.

\subsection{Case Study: Allocator Optimisation for \gem{}}

Allocator optimisation in realistic workloads is challenging: exploration is expensive, and results must align with system behaviour in the wild. 
We investigate the potential of memory allocator optimisation through a case study on \gem{}, a large and complex system simulator.

\vspace{0.1 cm}
\noindent\textit{Synthetic Benchmarking with \randmalloc{}. }%
Direct optimisation against the full \gem{} system would be prohibitively expensive, as a single run may take hours or days. We therefore employ \randmalloc{}~\cite{Langdon:2025:UKCI} as a proxy benchmark to effectively explore the allocator parameter space. 
\randmalloc{} provides a synthetic workload, generated from a seed trace of the real system, to exercise the memory allocation behaviour while remaining representative of the memory behaviour observed in \gem{}. This can also be used to measure and emulate the workload patterns of other systems, in order to provide a proxy to optimise over.

\vspace{0.1 cm}
\noindent\textit{Evaluation on \gem{}. }%
To assess the effectiveness of our approach, we transfer the best-performing parameter configurations to \gem{}.
We evaluated \gem{}'s system emulation~(SE) mode. 
For the SE mode, we rely on benchmark programs from previous work on fuzz testing of \gem{}~\cite{searchsysASE2025,searchsysICSE2025}.



\section{Evaluation}
\label{sec:evaluation}
We evaluated \tool{} on \gem{} using 50 C test programs from \searchsys{}’s datasets \cite{searchsysASE2025,searchsysICSE2025}. The study analyses \tool's impact on peak and average heap size, memory release rate, and instruction counts under four configurations:
\begin{itemize}[nosep,noitemsep,leftmargin=11pt,itemsep=0pt, topsep=0pt]
\item\textbf{(1) Def \texttt{glibc}:} unmodified \gnumem (baseline)
\item\textbf{(2) Opt \texttt{glibc}:} \gnumem tuned with \tool{}
\item\textbf{(3) Def \texttt{tcmalloc}:} unmodified \googlemem (baseline)
\item\textbf{(4) Opt \texttt{tcmalloc}:} \googlemem tuned with \tool{}
\end{itemize}
We then ask the following research questions (RQ):
\begin{list}{}{\leftmargin=0.5em
                \rightmargin=0.7em
                \topsep=0pt
                \partopsep=0pt
                \parsep=0pt
                \itemsep=0pt}
\item \textbf{RQ1:} \textit{What are the trade-offs between runtime performance and memory efficiency along the Pareto front of configurations discovered by \tool{}?}
\item \textbf{RQ2:} \textit{How effective is \tool{}'s optimisation in reducing memory consumption and execution time in industrial simulation workloads?}
\end{list}

\vspace{0.1cm}
\noindent For each case study, we compiled our generated C test inputs for use in \gem{}’s SE mode, executed them under the four configurations, and measured memory usage with Valgrind~\cite{valgrind} and instruction counts with \href{https://perfwiki.github.io/main}{perf}. We use a population size of 24, and run for 500 generations before stopping across all repetitions.
We used Valgrind~3.18.1 and perf~5.15.184, running on an Intel Xeon D-1548 (2.0\,GHz, 8~cores) with 64\,GB RAM, 8\,GB swap, and Ubuntu 22.04.5 LTS (x86\_64).


\begin{figure} [t!]
    \includegraphics[width=1.02\linewidth,clip]{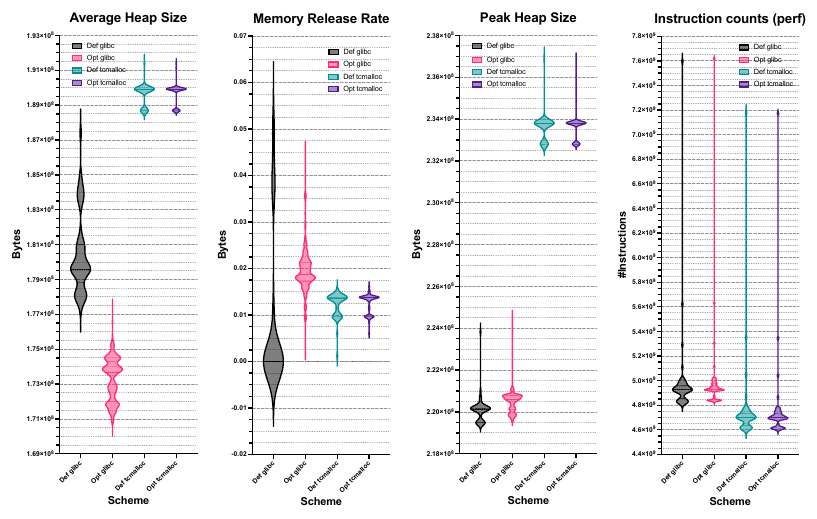}
    \vspace{-0.6cm}
    \hspace{-0.5cm}
    \caption{Comparison of default and \tool{}-optimised configurations of \gnumem (\texttt{glibc}) and \googlemem (\texttt{tcmalloc}). From left to right: average heap size, Memory release Rate, peak heap size, and instruction counts, as measured with Valgrind, perf, and \gem{}. Values are all pareto optimal values.}
    \vspace{-0.8cm}
    \label{fig:res}
\end{figure}

\noindent\textbf{Results: RQ1.} The hypervolume and Pareto front reveal trade-offs between the two allocators. For glibc, we achieve a mean hypervolume of $9.37 \times 10^{16}$. The Pareto fronts average 3 solutions per run, across 0.095\% of the instruction space and 0.162\% of the peak-heap space relative to the best-performing configuration on the front, with a trade-off slope of $-0.216$. While TCMalloc achieves a mean hypervolume of $5.46 \times 10^{16}$, with smaller Pareto fronts averaging 1.6 solutions per run. While its instruction span is constrained to just 0.005\%, it offers greater potential with a 0.259\% peak-heap span; however, it also has a much steeper trade-off of $-3.17$. 
This demonstrates glibc malloc's default configuration allows for more gradual trade-offs, while TCMalloc operates closer to optimal performance boundaries, constraining the search space but requiring more aggressive trade-offs between objectives. Full RQ1 results are provided in the zip file at \cite{zenodo:green}.

\noindent\textbf{Results: RQ2. }%
For \textbf{average heap usage}, \texttt{glibc} shows a clear improvement: Tuning reduced the mean from 180\,428\,315 to 173\,293\,862, with a tighter standard deviation ($2.2$M to $1.6$M). This indicates both greener execution -- due to a lower memory footprint -- and greater performance stability. For \texttt{tcmalloc}, the mean remains nearly unchanged, though the density distribution in \autoref{fig:res} shows less variance around hotspots, improving predictability.
Neither allocator shows reductions in \textbf{peak heap usage}. \texttt{glibc}’s average slightly increases (220\,104\,049 to 220\,523\,190), while \texttt{tcmalloc} remains stable. This shows \textbf{peak heap usage} is close to its minimum.
\texttt{glibc} benefits significantly in \textbf{memory release rate}: The free rate rises from $0.0080$ to $0.0196$, reducing retention. \texttt{tcmalloc} also improves, with both average release and minimum release increasing.
For \textbf{instructions executed}, \texttt{glibc} improves from $4.992 \times 10^9$ to $4.990 \times 10^9$, with standard deviation tightened. \texttt{tcmalloc} shows a clearer benefit: reducing average instructions from $4.77 \times 10^9$ to $4.76 \times 10^9$, with variance shrinking as well. These reductions translate into greener execution and improved efficiency.

In best-case runs, we found a single instance of \googlemem achieved a significant improvement over the unoptimised baseline: specifically, a 4.65\% reduction in instruction count as well as a simultaneous 2.06\% reduction in peak heap usage. In contrast, the best case run for \texttt{glibc} demonstrated a tradeoff between the two optimisation parameters, though it achieved a larger improvement in each metric individually. This suggests that while the transferability from the synthetic benchmark is sufficient, it could benefit from further tuning.

Overall, allocator tuning reduces memory usage and instruction count in \texttt{glibc}, while improving consistency in both \texttt{glibc} and \texttt{tcmalloc}, demonstrating tangible gains in both sustainability and performance. The improvements found by the synthetic benchmark could effectively translate to real workloads. In \texttt{glibc}, the tuned parameters reduced average memory and instructions and increased release rate significantly -- matching synthetic predictions of greener, more efficient behaviour. In \texttt{tcmalloc}, while mean values remained largely stable, the improved consistency of both heap usage and instruction counts mirrors the synthetic outcomes. Full RQ2 results are provided in the zip file at \cite{zenodo:green}.

\section{Conclusion}

We introduced \tool{}, a search-based framework for memory allocator parameter optimisation using lightweight benchmarking. With \gem{} we observed reductions in heap usage and instruction counts for both \textsc{glibc} and \mbox{\googlemem{}}, highlighting malloc parameter optimisation as a practical approach for efficient and greener systems. The combination of search-based optimisation with lightweight benchmarking opens the door to investigating other aspects of complex software using this strategy, such as \gem{}’s full system (FS) mode, and broader targets including VMs, simulators, emulators, and interpreters.


\bibliographystyle{splncs04}
\bibliography{references}

\end{document}